\documentclass[conference]{IEEEtran}
\IEEEoverridecommandlockouts
\usepackage[T1]{fontenc}
\usepackage{cite}
\usepackage{amsmath,amssymb,amsfonts}
\usepackage{algorithmic}
\usepackage{algorithm}
\usepackage{graphicx}
\ifCLASSOPTIONcompsoc
 \usepackage[caption=false,font=normalsize,labelfont=sf,textfont=sf]{subfig}
\else
 \usepackage[caption=false,font=footnotesize]{subfig}
\fi
\usepackage{eso-pic}
\usepackage{textcomp}
\usepackage{verbatim}
\usepackage{graphicx}
\usepackage{booktabs}
\usepackage{xcolor}
\def\BibTeX{{\rm B\kern-.05em{\sc i\kern-.025em b}\kern-.08em
    T\kern-.1667em\lower.7ex\hbox{E}\kern-.125emX}}

\begin{document}

\AddToShipoutPictureBG*{
    \AtTextLowerLeft{
        \put(0,-30){
            \parbox{\columnwidth}{
                \scriptsize
                \copyright~2024 IEEE. Personal use of this material is permitted. Permission from IEEE must be obtained for all other uses, in any current or future media, including reprinting/republishing this material for advertising or promotional purposes, creating new collective works, for resale or redistribution to servers or lists, or reuse of any copyrighted component of this work in other works.
            }
        }
    }
}
\title{Low-Complexity Soft-Feedback Detector for AFDM Systems\\
}

\author{\IEEEauthorblockN{Taohe Chen\IEEEauthorrefmark{1}, Yin Xu\IEEEauthorrefmark{1}, Tianyao Ma\IEEEauthorrefmark{1}, Aimin Tang\IEEEauthorrefmark{1}, Qu Luo\IEEEauthorrefmark{2}, Dazhi He\IEEEauthorrefmark{1}, Wenjun Zhang\IEEEauthorrefmark{1}}
\IEEEauthorblockA{\IEEEauthorrefmark{1} Shanghai Jiao Tong University,
Shanghai, China\\
\IEEEauthorrefmark{2} University of Surrey, Guildford, U.K.\\
Email: {\{taohe.chen, xuyin, mty0710, tangaiming\}@sjtu.edu.cn,\{q.u.luo\}@surrey.ac.uk,\{hedazhi, zhangwenjun\}@sjtu.edu.cn}
}
\thanks{This work was supported in part by the National Natural Science Foundation of China Program under Grant (62371291, 6242211) and Shanghai Municipal Science and Technology Commission Project (Grant No. 25DP1500200).}

\thanks{The corresponding author is Yin Xu (e-mail: xuyin@sjtu.edu.cn).}
}

\maketitle

\begin{abstract} 
Affine frequency division multiplexing (AFDM), an emerging multi-carrier modulation scheme, has garnered significant attention due to its resilience to Doppler shifts and capability to achieve full diversity in doubly dispersive channels. However, existing data detection algorithms for AFDM systems face a significant trade-off between computational complexity and accuracy. In this paper, a novel low-complexity data detection scheme, termed the soft-feedback detector (SFD), is proposed. Particularly, building upon a maximum ratio combining (MRC) estimator framework, the SFD leverages the \textit{a priori} symbol distribution to mitigate error propagation during iterative detection. Specifically, soft-decision feedback is incorporated as extrinsic information derived from the log-likelihood ratios of the transmitted symbols. As a result, the proposed detector significantly enhances detection accuracy while maintaining low computational complexity. Simulation results demonstrate that the SFD consistently outperforms benchmark decision-feedback detectors. In particular, compared with the conventional MRC detector, the proposed scheme achieves approximately a 3 dB signal-to-noise ratio (SNR) gain at the bit error rate (BER) of $10^{-3}$. 
\end{abstract}


\section{Introduction} 
The evolution towards 6G envisions ubiquitous connectivity across high-mobility scenarios, such as high-speed railways and non-terrestrial networks \cite{alliance2024itu}. However, high mobility introduces severe Doppler shifts, resulting in time-varying doubly dispersive channels that challenge the orthogonality of traditional multi-carrier waveforms such as the orthogonal frequency division multiplexing (OFDM) waveform. To this end, affine frequency division multiplexing (AFDM) has been proposed as a compelling waveform candidate in recent years \cite{bemani2023affine}. By virtue of its chirp-based subcarrier structure, AFDM can achieve full diversity gain in doubly dispersive channels, rendering it exceptionally robust for high-mobility communications.

While AFDM achieves full diversity in doubly dispersive channels, it inevitably encounters severe inter-symbol interference (ISI). Conventional linear equalizers, such as the minimum mean square error (MMSE) equalizer, rely on matrix inversion to mitigate the ISI. However, unlike the diagonal structure in OFDM, the effective channel matrix for AFDM is banded. Consequently, standard MMSE algorithms incur prohibitive computational complexity, motivating the design of novel detection schemes tailored for banded AFDM channels. 

Several AFDM data detection algorithms have recently been proposed in the literature. In \cite{bemani2022low}, an improved linear minimum mean square error (LMMSE) equalization and a maximum ratio combining (MRC) based decision feedback equalization (DFE) scheme (MRC-DFE) are proposed to reduce the computational complexity of traditional algorithms. In \cite{wu2024afdm}, the message passing (MP) algorithm is applied for AFDM data detection, utilizing the inherent channel sparsity for efficient interference cancellation. Additionally, a Turbo equalization scheme is introduced for Polar-coded AFDM based on LMMSE and belief propagation (BP) decoding in \cite{qu2024turbo}. However, existing data detection algorithms face a tradeoff between accuracy and complexity. Although the MRC-DFE exhibits a computationally efficient structure, it fails to exploit the \textit{a priori} distribution of transmitted symbols during iterative feedback. Instead, it directly utilizes the estimated symbols from each iteration, leading to error propagation. In contrast, the Turbo detector and MP detector introduce a soft-information passing mechanism, achieving superior BER performance, but incur high computational complexity. Therefore, how to achieve efficient data detection for AFDM systems with low complexity is still an open problem.\IEEEpubidadjcol

To this end, a novel data detection scheme is proposed for AFDM systems in this paper. The key mechanism of our proposed scheme is to introduce a soft-decision feedback to the conventional decision-feedback equalizer. Thus, it is named the soft-feedback detector (SFD) in the remainder of this paper. More specifically, the SFD leverages the \textit{a priori} distribution of transmitted symbols during iterative feedback. Furthermore, by simplifying the calculation of the estimated variance during soft feedback, the proposed scheme eliminates the need for matrix inversion to reduce the computational complexity. Thanks to the introduced soft-decision feedback information, the propagation error in the algorithm iteration can be highly reduced. Consequently, the data detection accuracy is highly improved compared with the existing equalization algorithms. 
The effectiveness of our proposed design is validated by simulation results, which demonstrate that our proposed design can accelerate the convergence of data detection, while achieving a lower mean squared error (MSE) compared to the benchmark MRC-DFE. In terms of the bit error rate (BER) metric, the SFD can achieve an approximately 3 dB signal-to-noise ratio (SNR) gain at the BER of $10^{-3}$ compared to the MRC-DFE. In addition, the SFD achieves the lowest total computational complexity in terms of floating-point operations (FLOPs) among the evaluated iterative detectors.

\emph{Notation}: For a vector $\mathbf{x}$, $x[i]$ denotes its $i$-th element, and $\|\mathbf{x}\|$ denotes its Euclidean norm. 
For a matrix $\mathbf{X}$, $X[i,j]$ denotes the $(i,j)$-th entry. The superscripts $(\cdot)^{-1}$, $(\cdot)^H$ denote the inverse and conjugate transpose operators, respectively. 
$\mathbb{C}^{M \times N}$ represents the space of $M \times N$ complex matrices. Furthermore, the superscript $(\cdot)^{(t)}$ denotes the iteration index.


\section{Signal Model}
Let $\mathbf{x} \in \mathbb{A}^{N \times1}$ denote the vector of the transmitted symbols in the discrete affine Fourier transform (DAFT) domain, where $\mathbb{A}$ represents the quadrature amplitude modulation (QAM) alphabet. The DAFT domain symbols will be transformed into the time domain by using inverse DAFT (IDAFT), expressed as
\begin{equation}
    s[n]=\frac{1}{\sqrt{N}}\sum^{N-1}_{m=0} {x[m]e^{\imath 2 \pi \left(c_1 n^2+c_2 m^2+ \frac{nm}{N}\right)}},
\end{equation}
where $n = 0,...,N-1$, $c_1 = \frac{2\left(\alpha_{\mathrm{max}}+\xi_{\nu}\right)+1}{2N}$ and $c_2$ is any irrational number \cite{bemani2023affine}. $\xi_{\nu}$ is a guard parameter decided by the fractional Doppler interference. $\alpha_{\mathrm{max}}$ is the normalized maximum Doppler shift of the current doubly dispersive channel. The matrix form of AFDM modulation takes the form
\begin{equation}
    \mathbf{s} = \mathbf{A}^{-1}\mathbf{x} = \mathbf{\Lambda}_{c_1}^H \mathbf{F}^H \mathbf{\Lambda}_{c_2}^H \mathbf{x},
\end{equation}
where $\mathbf{A}=\mathbf{\Lambda}_{c_1} \mathbf{F} \mathbf{\Lambda}_{c_2}$ denotes the DAFT matrix, which is a unitary matrix, $\mathbf{\Lambda}_{c_{\mu}} = \mathrm{diag} \left(e^{-j2\pi c_{\mu} n^2}\right), n=0,...,N-1$ and $\mathbf{F}$ is the discrete Fourier transform (DFT) matrix.

In order to combat multipath propagation, a chirp-periodic prefix (CPP) is used in AFDM time domain symbols, which is given by
\begin{equation}
    s[k] = s[N+k]e^{-j2\pi c_1(N^2 + 2Nk)},\quad k=-L_{\mathrm{cpp}},...,-1
\end{equation}
where $L_{\mathrm{cpp}}$ is the length of CPP.

By virtue of the delays and the Doppler shifts of dual dispersion channel paths, the received symbols in the time domain can be expressed as
\begin{equation}
    r[n]=\sum^{\infty}_{l=0}s[n-l]g_{n}[l]+w[n],\quad n = 0,...,N-1,
\end{equation}
where $w[n]$ denotes the additive white Gaussian noise (AWGN) with a vector form of $\mathbf{w} \sim \mathcal{CN}(0,\mathbf{N_0 I})$. The impulse response of the channel at time $n$ and delay $l$ is
\begin{equation}
    g_{n}[l]=\sum_{i=1}^{P}h_{i}e^{-\jmath 2 \pi f_i n}\delta(l-l_{i}),
\end{equation}
where $\delta(\cdot)$ is the Dirac delta function. The channel consists of $P$ paths, where the $i$-th path is characterized by a complex gain $h_i$, an integer delay $l_i$ in samples, and a Doppler shift $f_i$. The normalized Doppler shift of the $i$-th path is defined as $\alpha_i = N f_i \in [-\alpha_{\mathrm{max}}, \alpha_{\mathrm{max}}]$.

The demodulation operation takes the form
\begin{equation}
    y[m]=\frac{1}{N}\sum^{N-1}_{n=0}{r[n]e^{-\jmath 2 \pi \left(c_1 n^2+c_2 m^2+ \frac{nm}{N}\right)}},
\end{equation}
where $m = 0,...,N-1$. The vector of DAFT domain output symbols $\mathbf{y}$ is given by
\begin{equation}
\mathbf{y} = \mathbf{H}_{\mathrm{eff}} \mathbf{x} +  \tilde{\mathbf{w}},
\end{equation}
where $\tilde{\mathbf{w}} = \mathbf{Aw}$ denotes the noise in the DAFT domain and  $\mathbf{H}_{\mathrm{eff}}$ represents the efficient channel matrix in the DAFT domain, which contains $P$ components. Its $i$-th component takes the form
\begin{equation}\label{eq:frac doppler Heff}
|H_i[r,c]| \propto \left|\frac{\sin(N\phi_{r,c,i})}{N\sin\phi_{r,c,i}}\right|,
\end{equation}
where $\phi_{r,c,i} = \pi(r-c-p_i-a_i)/N $, and $p_i = 2N c_1 l_i + \alpha_i$ is the position parameter of the $i$-th path. 

\begin{figure}[!t]
    \centering
    \includegraphics[width=0.7\linewidth]{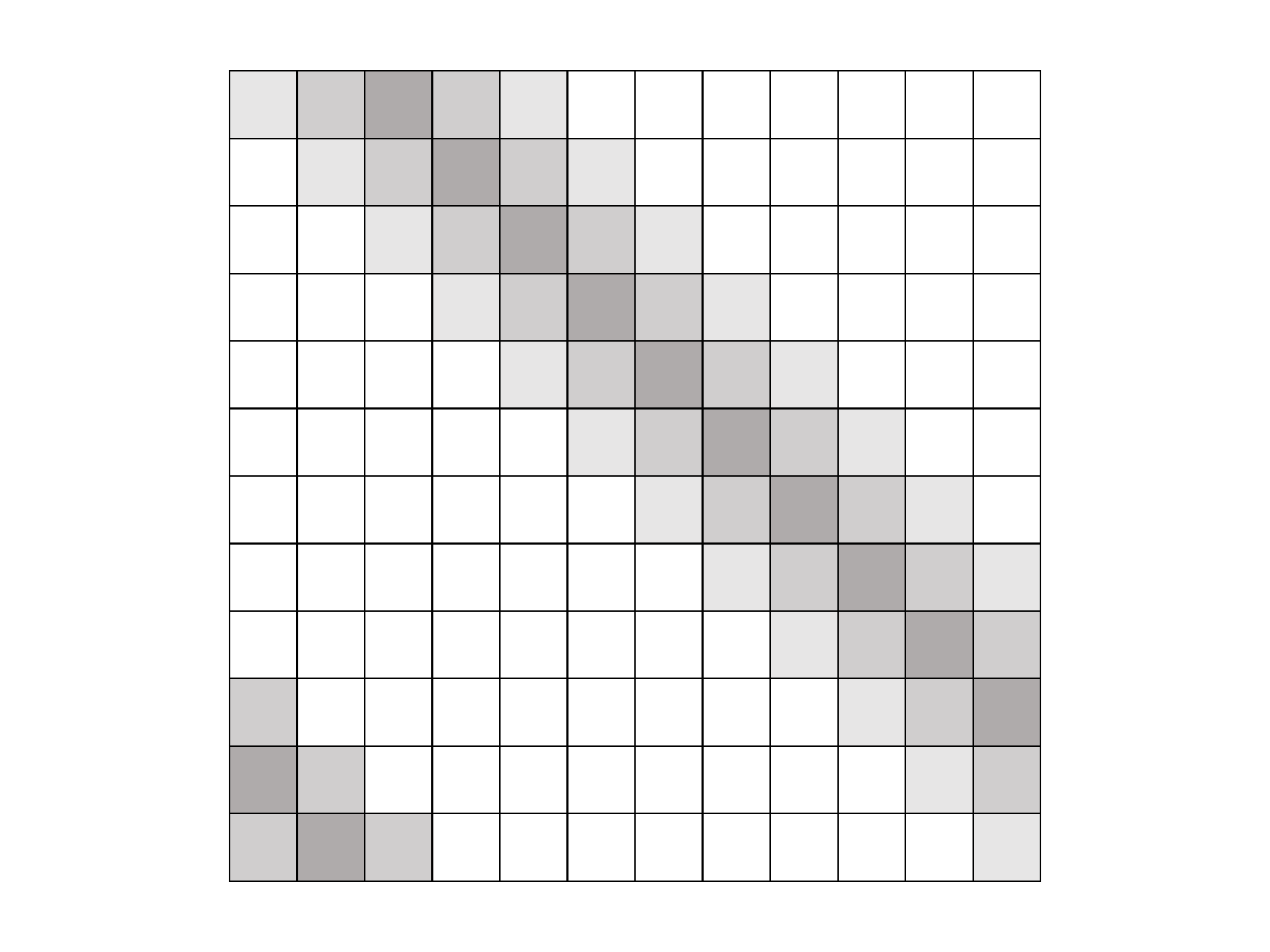}
    \caption{Illustration of the approximate band matrix structure. Gray cells represent non-zero entries, while white cells represent zero entries.}
    \vspace{-1em}
    \label{fig:Heff_B}
\end{figure}

As assumed in \cite{bemani2023affine}, each column of $\mathbf{H}_{\mathrm{eff}}$ contains $L$ non-zero entries under fractional Doppler shifts, where $L=\left( 2\xi_{\nu}+1\right)P$, as shown in Fig. \ref{fig:Heff_B}. Let $\mathcal{J}_c = \{r_1, \dots, r_L\}$ be the row index set of the non-zero entries in the $c$-th column of $\mathbf{H}_{\mathrm{eff}}$. Dually, let $\mathcal{I}_r = \{c_1, \dots, c_L\}$ denote the column index set of the non-zero entries in the $r$-th row.


\section{Detection Algorithm}
This paper focuses on the data detection for AFDM systems, i.e., given the received symbol $\mathbf{y}$ and the (estimated) effective channel $\mathbf{H}_{\mathrm{eff}}$, we aim to recover the transmitted symbol $\mathbf{x}$. The basic structure of our proposed SFD detector follows the maximum ratio combining estimator, incorporating a novel soft-decision feedback.  

\subsection{Maximum Ratio Combining Estimator}
To recover the transmitted symbols, our proposed SFD algorithm first constructs a low-complexity MRC estimator. The MRC estimator outputs estimated DAFT-domain symbols in each iteration by leveraging the \textit{a priori} information given by the last iteration. Subsequently, the output of the estimator is processed to derive \textit{a posteriori} information, which serves as the updated\textit{ a priori} knowledge for the next iteration. More details are elaborated as follows.

First, to efficiently eliminate ISI, we define a global residual at the $t$-th iteration as
\begin{equation}
    \Delta y^{(t)}[r] = y[r]-\sum_{c\in \mathcal{I}_r} H_\mathrm{eff}[r,c] \cdot \hat{x}^{(t-1)}[c],
\end{equation}
where $\hat{x}^{(t-1)}[c]$ is the estimation of $x[c]$ from the $(t-1)$-th iteration. 

In the conventional MRC-DFE, the contribution of the $c$-th transmitted symbol $x[c]$ to the $r$-th received symbol $y[r]$ at the $t$-th iteration can be expressed as
\begin{equation}
    \kappa ^{(t)}_{r,c}=\Delta y^{(t)}[r]+H_\mathrm{eff}[r,c] \cdot \hat{x}^{(t-1)}[c].
\end{equation}
However, this hard feedback scheme implicitly treats all symbols with equal confidence, inevitably resulting in error propagation. To mitigate this, our approach utilizes soft information for feedback, modifying the contribution function as
\begin{equation}
    u^{(t)}_{r,c}=\Delta y^{(t)}[r]+H_\mathrm{eff}[r,c] \cdot E^{(t-1)}_{\mathrm{sym}}[c],
\end{equation}
where $E^{(t-1)}_{\mathrm{sym}}[c]$ denotes the conditional soft symbol expectation of $x[c]$ which is fed back from the previous iteration. Its specific definition will be detailed in the next subsection.

Leveraging the principle of MRC, the combined soft decision statistic can be formulated as
\begin{equation}
    \begin{aligned}
         g[c]^{(t)} &= \sum_{r \in \mathcal{J}_c} H_{\mathrm{eff}}^{*}[r,c] \cdot E^{(t-1)}_{\mathrm{sym}}[c] \\
         &=\sum_{r \in \mathcal{J}_c} H_{\mathrm{eff}}^{*}[r, c] \Delta y^{(t)}[r] + d_c \cdot E^{(t-1)}_{\mathrm{sym}}[c],
    \end{aligned}
\end{equation}
where $d_c = \sum_{r \in \mathcal{J}_c} \left|H_{\mathrm{eff}}[r,c]\right|^2$. Since $\left| \mathbf{H}_{\mathrm{eff}}\right|$ is an approximate circulant matrix, the parameter $d_c$ can be calculated only once in the whole detection to reduce complexity \cite{bemani2023affine}, so that $d_c=d$ for $c=0,\dots, N-1$.

Based on the derivation in \cite{bemani2022low}, the estimated DAFT domain symbols are given by 
\begin{equation}
    \mathbf{\hat{x}}^{(t)} = \frac{\mathbf{g}^{(t)}}{d+\gamma^{-1}},
\end{equation}
where $\gamma$ represents the SNR. 

The estimation is followed by the iteration stopping judgment. We employ a relative convergence criterion with a tolerance of $T_{\mathrm{error}}$. 
Consequently, the residual vector is updated to reflect the new estimate as follows:
\begin{equation}
    \Delta y^{(t)}[r]  = \Delta y^{(t-1)}[r]  - H_{\mathrm{eff}}[r,c]\left( \hat{x}^{(t)}[c]-\hat{x}^{(t-1)}[c]\right),
    \label{eq:residual updating}
\end{equation}
where $r \in \mathcal{J}_c$ for $c=0,\dots, N-1$.

\subsection{Soft Decision Feedback}

\begin{figure}[!t]
    \centering
    \includegraphics[width=0.7\linewidth]{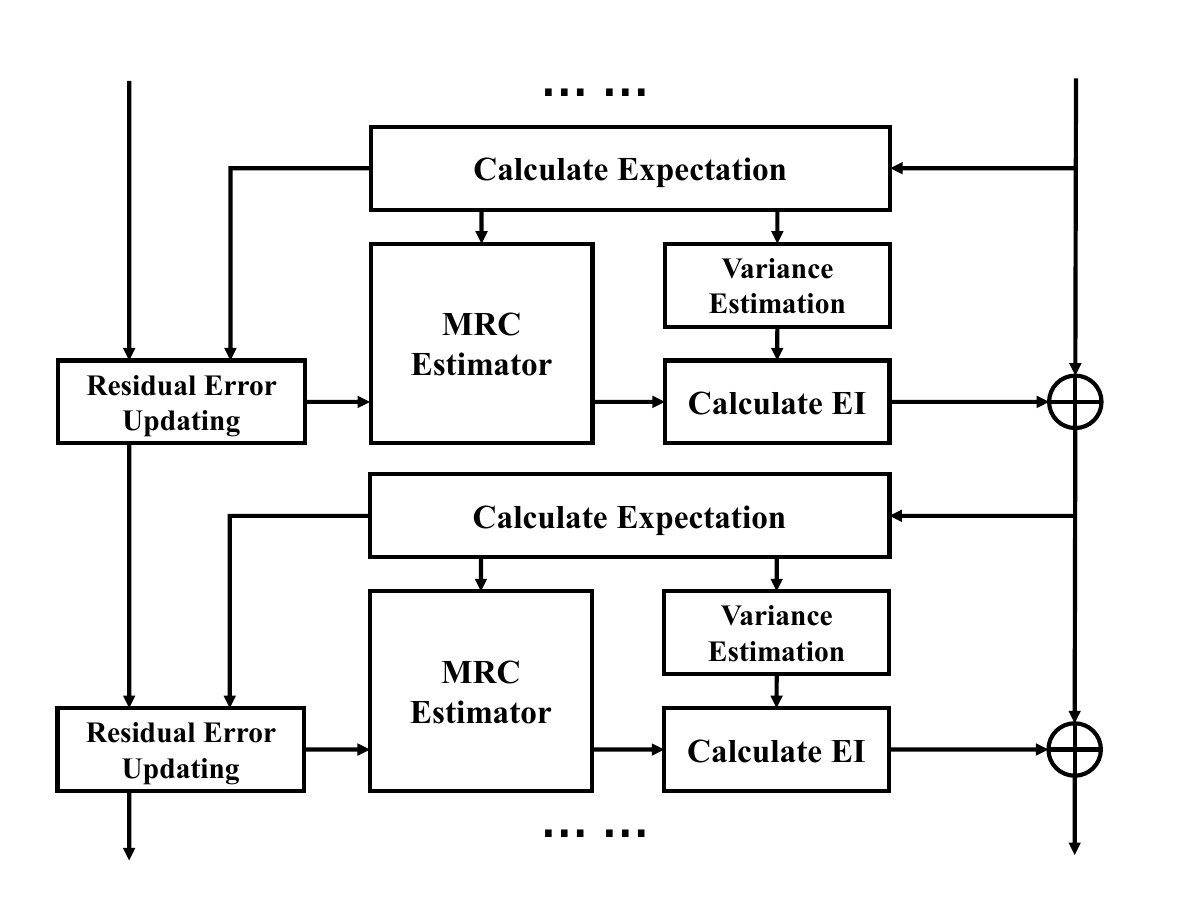}
    \caption{The iteration structure of SFD.}
    \label{fig:iterationStructure}
\end{figure}
Although the MRC estimator achieves lower complexity than the classical MMSE algorithm, its performance highly depends on the error propagation in each iteration. To this end, we propose a soft-decision feedback mechanism in the MRC estimator to improve the data detection accuracy. 
To achieve information accumulation and exploitation across iterations, we proceed to derive the \textit{a priori} information for the symbol estimation in each iteration to reduce the error propagation, which leverages the \textit{a priori} distribution of transmitted symbols during iterative feedback. Without loss of generality, the QPSK constellation is adopted to present the design. However, the proposed mechanism can be readily extended to higher-order modulation. We use QPSK for simplicity of the presentation.

Define $x_\mathrm{bit}[c,b]$ as the $b$-th bit ($b = 1,2$) in the $c$-th QPSK symbol and $\hat{x}^{(t)}_\mathrm{bit}[c,b]$ as the $b$-th bit ($b = 1,2$) in the $c$-th estimated information symbol. Let $z_{c,b}\in \{0,1\}$ be the actual value of $x[c,b]$. After the MRC estimator, the output estimated symbols are utilized to calculate extrinsic information (EI), defined in the form of log-likelihood ratio (LLR), expressed as
\begin{equation}
    L^{(t)}_{\mathrm{ext}}[c,b] = \ln{ \frac{\Pr \left( \hat{x}^{(t)}_\mathrm{bit}[c,b]|x_\mathrm{bit}[c,b] = +1\right)}{\Pr\left(\hat{x}^{(t)}_\mathrm{bit}[c,b]|x_\mathrm{bit}[c,b] = -1\right)}},
    \label{eq:extrinsic information ori}
\end{equation}
where $c = 0,\dots, N-1$ and $b=1,2$.

The probability density function (PDF) of $\hat{x}^{(t)}_\mathrm{bit}[c,b]$ is extensively approximated as Gaussian, which is expressed as 
\begin{equation}
    \begin{aligned}
         & f \left(\hat{x}^{(t)}_\mathrm{bit}[c,b] \,\big|\, x_\mathrm{bit}[c,b] = z_{c,b} \right)\\
         & = \frac{1}{\sqrt{2\pi}\sigma^{(t)}_\mathrm{bit}[c,b]} \exp \left( -\frac{ \left( \hat{x}^{(t)}_\mathrm{bit}[c,b] - z_{c,b} \right)^2 }{2\sigma^{2,(t)}_\mathrm{bit}[c,b]} \right),
    \end{aligned}
    \label{eq:Gaussian_distribution}
\end{equation}
where the parameters $\sigma^{(t)}_\mathrm{bit}[c,b]$, for$ c=0,\dots, N-1$ and $b=1,2$, represent the standard deviations of the corresponding estimated bits.

Substituting \eqref{eq:Gaussian_distribution} into \eqref{eq:extrinsic information ori} yields
\begin{equation}
    L^{(t)}_{\mathrm{ext}}[c,b]  =\frac{\sqrt{2}}{\sigma^{2,(t-1)}_\mathrm{bit}[c,b]}\hat{x}^{(t)}_\mathrm{bit}[c,b].
\end{equation}

In the SFD, the extrinsic information acts as an incremental update to the LLRs. Specifically, the \textit{a posteriori} LLR from the $(t-1)$-th iteration is utilized as the \textit{a priori} LLR for the $t$-th iteration, given by
\begin{equation}
    L_{\mathrm{Pri}}^{(t+1)}[c,b] = L_{\mathrm{Post}}^{(t)}[c,b]
\end{equation}
where
\begin{equation}
    \begin{aligned}
        L_{\mathrm{Post}}^{(t)}[c,b]  &= \ln{\frac{\Pr \left( x[c,b] = +1|\hat{x}^{(t-1)}[c,b]\right)}{\Pr \left(x[c,b] = -1 | \hat{x}^{(t-1)}[c,b]\right)}}\\
        & = L^{(t)}_{\mathrm{ext}}[c,b] + L_{\mathrm{Pri}}^{(t)}[c,b].
    \end{aligned}
\end{equation}

\begin{table}[t]
\caption{Comparison of Computational Complexity}
\begin{center}
\renewcommand{\arraystretch}{1.3}
\renewcommand{\tabcolsep}{1mm}
\begin{tabular}{l c c}
\hline
\textbf{Algorithm} & \textbf{Complexity} & \textbf{Approximate FLOPs} \\
\hline
MMSE & $\mathcal{O}(N^3)$ & $24N^3$ \\
MF-MP & $\mathcal{O}(NL^2 + N_{\mathrm{iter}}NL)$ & $32NL^2 + N_{\mathrm{iter}}(32NL + 120N)$ \\
MRC-DFE & $\mathcal{O}(N_{\mathrm{iter}}NL)$ & $N_{\mathrm{iter}} \cdot N(16L + 17)$ \\
SFD (Prop.) & $\mathcal{O}(N_{\mathrm{iter}}NL)$ & $N_{\mathrm{iter}} \cdot N(16L + 51)$ \\
\hline
\multicolumn{3}{l}{\footnotesize $N$: Subcarriers, $L$:Non-zero bandwidth, $N_{\mathrm{iter}}$: Iterations.} \\
\end{tabular}
\label{tab:complexity}
\end{center}
\end{table}

\addtolength{\topmargin}{0.08in}

Leveraging the \textit{a priori} information passed from the previous iteration, the SFD computes the soft symbol expectation for the $t$-th iteration. Under the QPSK constellation, the expectation of $x[c]$ is expressed as
\begin{equation}
    \begin{aligned}
         E^{(t)}_{\mathrm{sym}}[c] & = \mathrm{E}[x[c]|\hat{x}[c]^{(t)}] \\
         & =\frac{\tanh{\left( L_{\mathrm{Post}}^{(t)}[c,1] \right)} + j\tanh{\left( L_{\mathrm{Post}}^{(t)}[c,2] \right)}}{\sqrt{2}}.
    \end{aligned}
    \label{eq:expectation}
\end{equation}
During the initialization phase, $E^{(t)}_{\mathrm{sym}}[c]=0$. The considered variance of $x[c]$ is given by \eqref{eq:variance}, as discussed in \cite{tuchler2002minimum}.
\begin{equation}
    v[c]^{(t)} = 1 - \left|E^{(t)}_{\mathrm{sym}}[c]\right|^2.
    \label{eq:variance}
\end{equation}

According to the derivation in \cite{fang2008low}, the variance of the estimated bit can be expressed as
\begin{equation}
    \begin{aligned}
        \sigma^{2,(t)}_\mathrm{bit}[c,b] &= \mathrm{Cov} \left(\hat{x}^{(t)}[c,b], \hat{x}^{(t)}[c,b] \mid x[c,b] = z_{c,b} \right) \\
        &=v[c]^{2,(t)} \theta^{(t)}_{c} \left( 1-v[c]^{(t)} \theta^{(t)}_{c} \right),
    \end{aligned}
\end{equation}
with
\begin{equation}
    \theta^{(t)}_{c} = H_{\mathrm{eff}}[c]^{H}\left(\mathbf{H}_{\mathrm{eff}} \mathbf{V}^{(t)} \mathbf{H}_{\mathrm{eff}}^H - \mathbf{R}_{\mathrm{noise}}\right) ^{-1}H_{\mathrm{eff}}[c],
\end{equation}
where $\mathbf{V}^{(t)} = \mathrm{diag} \left([v[1]^{(t)}, v[2]^{(t)},...,v[N]^{(t)}]\right)$, and $\mathbf{R}_{\mathrm{noise}}$ represents the DAFT domain noise covariance matrix. However, directly applying such a process would incur high computational costs due to the matrix inversion computation. Thus, we propose a low-complexity approach that approximately derives variance estimates through linear scaling of $v[c]^{(t)}$. As a result, the variance can be expressed as
\begin{equation}
    \sigma^{2,(t)}_\mathrm{bit}[c,b]  = \eta \left(1 - \left|E^{(t)}_{\mathrm{sym}}[c]\right|^2 \right),
\end{equation}
where $b=1,2$, and $\eta$ is an equivalent parameter. This simplification captures the reliability evolution during iterations without explicit matrix inversion computation. In the first iteration, $ \sigma^{2,(t)}_\mathrm{bit}[c,b] = 1$.

The detailed procedure of the SFD is shown in Algorithm \ref{alg:SFD} and the iterative structure is shown in Fig. \ref{fig:iterationStructure}.

\begin{algorithm}[!t]
\caption{SFD Detection for QPSK}\label{alg:SFD}
\begin{algorithmic}[1]
\REQUIRE{$\mathbf{H}_{\mathrm{eff}}$, $\mathbf{y}$, $\hat{\mathbf{x}}^{(0)}=\mathbf{0}$, $\mathbf{\Delta y}=\mathbf{y}$, $E^{(0)}_{\mathrm{sym}}[c]=0$, $\sigma^{2,(t)}_\mathrm{bit}[c,b] = 1$, $L_{\mathrm{Post}}^{(0)}[c,b]=0$, $c=0,\dots,N-1$,$b=1,2.$}
 \STATE Determine $L$ and $\mathcal{J}_c$.
 \STATE $d = \sum_{r \in \mathcal{J}_c} \left|H_{\mathrm{eff}}[r,c]\right|^2$
 \FOR{$t=1:t_{\mathrm{maxIter}}$}
  \FOR{$c=0:N-1$}
    \STATE $g[c]^{(t)} =\sum_{r \in \mathcal{J}_c} H_{\mathrm{eff}}^{*}[r, c] \Delta y^{(t-1)}[r] + d \cdot E^{(t-1)}_{\mathrm{sym}}[c]$
    \STATE $\mathbf{\hat{x}}^{(t)} = \frac{\mathbf{g}^{(t)}}{d+\gamma^{-1}}$
    \STATE $\Delta y^{(t)}[r] = $
    \STATE $ \Delta y^{(t-1)}[r] - H_{\mathrm{eff}}[r,c]\left( \hat{x}^{(t)}[c]-\hat{x}^{(t-1)}[c]\right)$
    \ENDFOR
    \IF{$\left| \left|\mathbf{\hat{x}}^{(t)}-\mathbf{\hat{x}}^{(t-1)}  \right| \right| \leq T_{error} \left| \left|\mathbf{\hat{x}}^{(t-1)}  \right| \right|$ }
     \STATE EXIT.
    \ENDIF
    \FOR{$c = 0: N-1$}
     \STATE $L^{(t)}_{\mathrm{ext}}[c,b]  =\frac{\sqrt{2}}{\sigma^{2,(t-1)}_\mathrm{bit}[c,b]}\hat{x}^{(t)}_\mathrm{bit}[c,b], \quad b=1,2$
     \STATE $ L_{\mathrm{Post}}^{(t)}[c,b]  = L^{(t)}_{\mathrm{ext}}[c,b] + L_{\mathrm{Pri}}^{(t)}[c,b]$
     \STATE $E^{(t)}_{\mathrm{sym}}[c] =\frac{\tanh{\left( L_{\mathrm{Post}}^{(t)}[c,1] \right)} + j\tanh{\left( L_{\mathrm{Post}}^{(t)}[c,2] \right)}}{\sqrt{2}}$
     \STATE $\sigma^{2,(t)}_\mathrm{bit}[c,b]  = \eta \left(1 - \left|E^{(t)}_{\mathrm{sym}}[c]\right|^2 \right)$
  \ENDFOR
 \ENDFOR
\end{algorithmic}
\end{algorithm}

\subsection{Complexity Analysis}
\label{subsec:complexity}
The computational complexity in terms of FLOPs is analyzed. A real addition or multiplication corresponds to  1 FLOP, while a complex multiplication and division correspond to 6 and approximately 15 FLOPs, respectively. Notably, the $\tanh$ and exponential functions are implemented via piecewise linear Look-Up Tables (LUT) \cite{xie2020twofold}, where fetching parameters for linear approximation involves one multiplication and one addition, equivalently costing approximately 5 FLOPs.

The complexity of SFD is primarily composed of two parts: the linear interference cancellation and the soft-information update. The former, involving sparse matrix-vector multiplications over an effective bandwidth $L$, incurs $16L$ FLOPs per subcarrier. The latter, which includes LLR computation and expectation estimation utilizing LUTs, introduces a constant overhead of approximately 51 FLOPs. Consequently, the total complexity of SFD is derived as:
\begin{equation}
    C_{\mathrm{SFD}} \approx N_{\mathrm{iter}} \cdot N (16L + 51),
\end{equation}
where $N_{\mathrm{iter}}$ denotes the number of iterations. 


The computational complexities and corresponding FLOPS for different detectors, including our proposed SFD, MMSE, MRC-DFE, and MF-MP, are summarized in Table \ref{tab:complexity}. We can see that the proposed SFD maintains the linear complexity $\mathcal{O}(N_{\mathrm{iter}}NL)$, as that of the MRC-DFE, since they are both designed under the MRC framework. 
In contrast, the MF-MP and MMSE incur significantly higher computational burdens of $\mathcal{O}(NL^2)$ and $\mathcal{O}(N^3)$, respectively.


\section{Simulation results}
\label{sec:simulation}
In this section, simulation results are provided to validate the effectiveness of the proposed SFD algorithm. Besides the MRC-DFE, the classical MMSE detector and the recently proposed MF-MP detector all serve as benchmark algorithms for comparison.

\begin{figure}[!t]
\centering
\subfloat[SNR = 8 dB.]{
    \includegraphics[width=0.45\columnwidth]{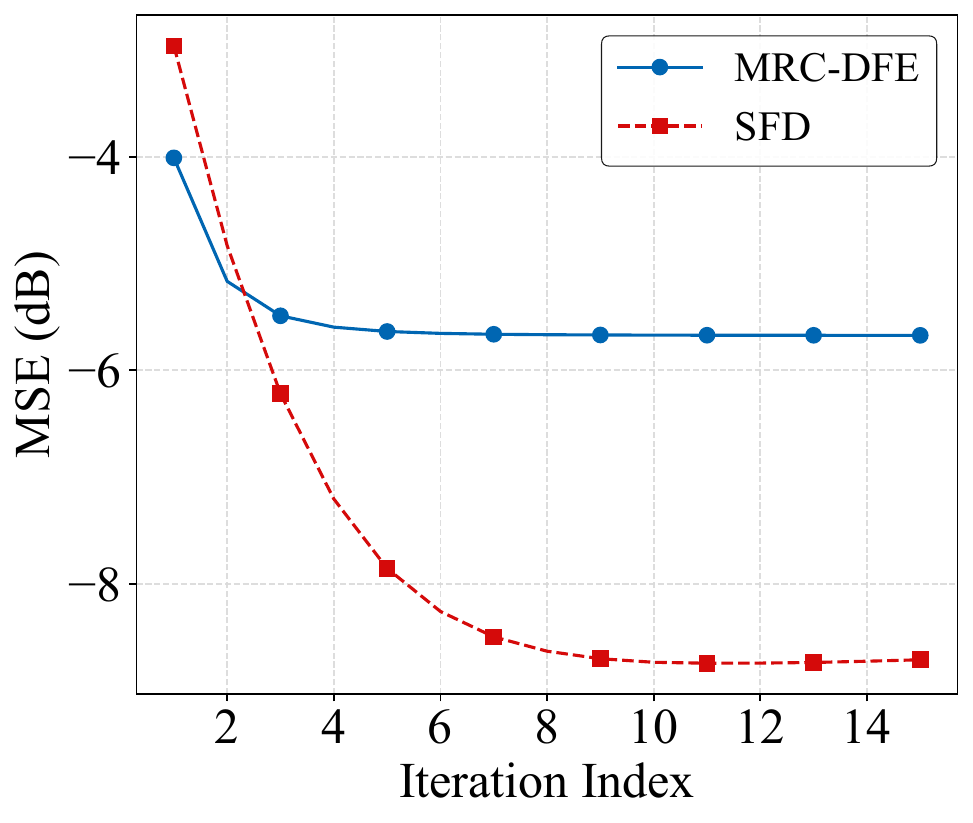}
    \label{fig:left}
}
\hfil 
\subfloat[SNR = 16 dB.]{
    \includegraphics[width=0.45\columnwidth]{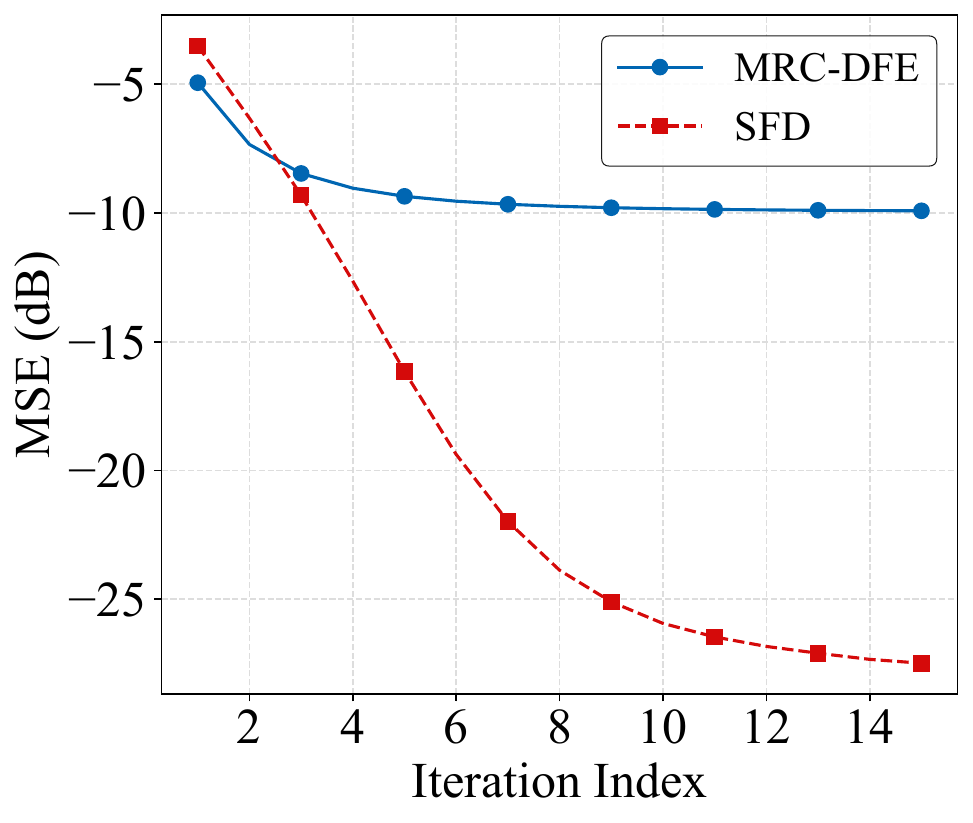}
    \label{fig:right}
}
\caption{Convergence performance comparison.}
\label{fig:convergence}
\end{figure}

First, the convergence behavior is evaluated by measuring the mean squared error (MSE) between the estimated and transmitted symbol vectors for the SFD and MRC-DFE, since they are under the same detector framework and with the lowest computational complexity as shown in Table \ref{tab:complexity}. The average MSE results, derived from 10,000 Monte Carlo trials at SNRs of 8 dB and 16 dB, are illustrated in Fig. \ref{fig:convergence}. As observed, the SFD utilizing soft information feedback exhibits significantly lower error floor compared to the MRC-DFE benchmark, confirming that soft information improves symbol estimation and effectively mitigates error propagation. Furthermore, beyond two iterations, the proposed design consistently attains a lower MSE, demonstrating superior performance at identical computational complexity.

\begin{figure}[t!]
    \centering
    \includegraphics[width=0.45\textwidth]{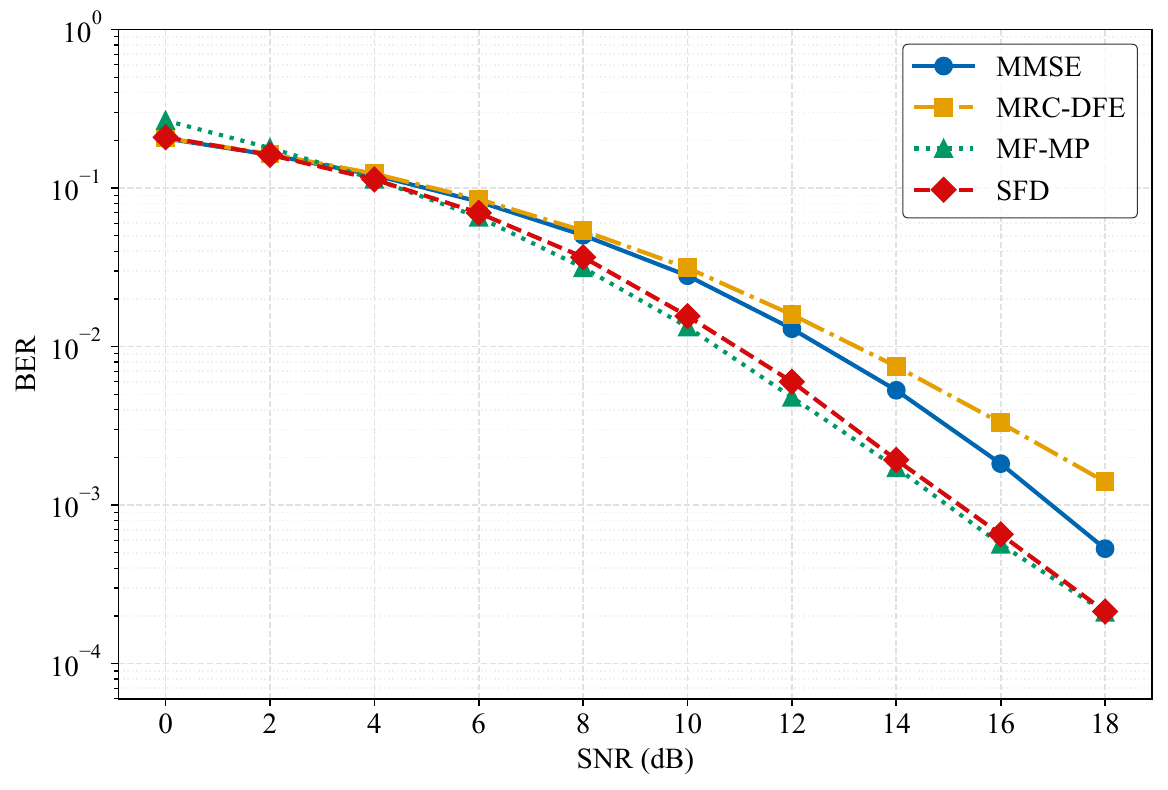} 
    \caption{BER performance comparison of different detectors.}
    \label{fig:BER}
\end{figure}

Next, the BER performance comparison is evaluated. The simulation employs a doubly dispersive channel model with $P=4$ paths, maximum delay $l_{\mathrm{max}}=3$, and maximum normalized Doppler shift $\alpha_{\mathrm{max}}=2$. The AFDM frame size is $N=512$. For iterative detectors, 10,000 Monte Carlo trials are performed. In this simulation, all the benchmarks are evaluated. To ensure a fair complexity comparison, a unified relative convergence threshold of $\epsilon = 0.01$ is adopted for both the proposed detector and the benchmarks.

The BER performance under different SNRs is illustrated in Fig. \ref{fig:BER}. The performance of MF-MP is better than MRC-DFE and MMSE. However, MF-MP incur higher computational complexity than MRC-DFE. This illustrates a trade-off between detection accuracy and computational complexity for the existing approaches. However, the proposed SFD achieves a superior trade-off between performance and complexity. Compared to the main benchmark MRC-DFE, the SFD provides substantial performance gains with a marginal increase in complexity. Compared to the more advanced MF-MP algorithm, the SFD algorithm performs comparably but has obviously lower complexity. Particularly, at the BER of $10^{-3}$, the SFD yields a SNR gain of approximately 3 dB compared to the MRC-DFE. This confirms that the soft feedback architecture effectively suppresses error propagation, so as to improve the detection accuracy.

\section{Conclusion}
A low-complexity data detector, termed the SFD, was developed for the AFDM systems. The proposed SFD adopted the low-complexity MRC framework. To improve the detection accuracy, a soft-decision feedback mechanism was designed through an extrinsic information transfer. The proposed approach can mitigate the error propagation inherent in conventional decision feedback detectors while maintaining low computational complexity. Simulation results demonstrated that the SFD algorithm can achieve superior detection accuracy compared to the existing benchmarks, yielding an SNR gain of approximately 3 dB over MRC-DFE at a BER of $10^{-3}$. 

The proposed design was validated in the single-input single-output (SISO) signal model. However, the current algorithm cannot be directly applicable to systems with multiple antennas, which is subject to future work.

\bibliography{IEEEabrv,citations}
\bibliographystyle{IEEEtran}

\end{document}